\documentclass[conference]{IEEEtran}
\IEEEoverridecommandlockouts
\usepackage{amsmath,amsfonts}
\usepackage{algorithmic}
\usepackage{algorithm}
\usepackage{array}
\usepackage[caption=false,font=normalsize,labelfont=sf,textfont=sf]{subfig}
\usepackage{textcomp}
\usepackage{stfloats}
\usepackage{url}
\usepackage{verbatim}
\usepackage{graphicx}
\usepackage{tabularx}
\usepackage{cite}
\usepackage{soul}
\usepackage{xcolor}
\usepackage{flushend}
\usepackage{nomencl}%nomenclature package
\usepackage{etoolbox} % For customizing the nomenclature list
\usepackage{ifthen} %used for nomenclature subsections
\usepackage{multirow}
\usepackage{placeins}
\usepackage{booktabs}
\usepackage{siunitx}
\usepackage{longtable}
\usepackage{bbm}
\usepackage{balance}

\newcommand{\reg}{\textit{REGULAR}}
\newcommand{\sch}{\textit{SCHEDULED}}
\newcommand{\lea}{\textit{LEAVE}}

\newcommand{\sto}{{\text{subject to:}}}
\renewcommand{\Pr}[1]{\text{Pr}\left(#1\right)}

\newcommand{\req}{{\text{req}}}
\newcommand{\E}{{\text{E}}}

\expandafter\def\expandafter\normalsize\expandafter{%
    \normalsize%
    \setlength\abovedisplayskip{-5pt}%
    \setlength\belowdisplayskip{6pt}%
    \setlength\abovedisplayshortskip{-5pt}%
    \setlength\belowdisplayshortskip{6pt}%
}

\renewcommand{\nomgroup}[1]{%
  \ifthenelse{\equal{#1}{V}}{%
    \vfil\penalty-10000 
    \item[\textit{Variables}]%
  }{%
    \ifthenelse{\equal{#1}{I}}{%
      \item[\textit{Indices/Sets}]}{%
    \ifthenelse{\equal{#1}{P}}{%
      \item[\textit{Parameters}]}{%
    }}%
  }%
}

\makenomenclature

\nomenclature[I1]{$\tau/\mathcal{T}$}{Time index of the monthly billing cycle.}
\nomenclature[I2]{$n$}{New user whose arrival triggers current optimization.}
\nomenclature[I3]{$i/\mathcal{I}$}{Set of all active users connected to the charging station, including user $n$, $\mathcal{I} = \mathcal{A}_\text{sch} \cup \mathcal{A}_\text{reg} \cup \{  n \}$.}
\nomenclature[I4]{$m$}{Charging service choice selected by user $n$. $m \in \{\reg{}, \sch{}, \lea{}\}$.}
\nomenclature[I5]{$i,j/\mathcal{A}_m$}{Set of existing users with service option $m$.}
\nomenclature[I6]{$T^\text{start}$}{Time index of current optimization, i.e. when user $n$ arrives at the station.}
\nomenclature[I7]{$T_i^{\text{end}}$}{The time index when user $i$ is no longer charging. This corresponds to the departure time indicated by a \sch{} user or the time index when a \reg{} user's battery is full.}
\nomenclature[I8]{$T^\text{end}$}{Time index of latest departure across all users $i \in \mathcal{I}$. $T^\text{end} = \max_{i \in \mathcal{I}}T_i^{\text{end}}$.}

\nomenclature[I9]{\textbf{}}{}  % Forces a blank line

\nomenclature[P1]{$\Delta t$}{Time step of the system, in hours.}
\nomenclature[P2]{$p^{\text{max}}$}{Maximum power of a charger, in kW.}
\nomenclature[P3]{$\eta$}{Charger efficiency.}
\nomenclature[P4]{$c_{\tau}$}{Utility rate for electricity at time index $\tau$, in \$/kWh.}
\nomenclature[P5]{$c_{d}$}{Utility rate for demand charge, in \$/kW.}
\nomenclature[P6]{$\E_i^\req$}{Required energy of user $i$, in kWh.}

\nomenclature[V1]{$p$}{Power profiles of all users in $\mathcal{I}$ to be optimized in current optimization, in kW.}
\nomenclature[V2]{$p^\text{sch}_{i,\tau}$}{Charging power for \sch{} user $i$ at time $\tau$ in the current optimization, in kW.}
\nomenclature[V3]{$e_{i, \tau}$}{Energy level for user $i$ at time index $\tau$, in kWh.}
\nomenclature[V4]{$G_\tau$}{Station charging load at time index $\tau$, in kW.}
\nomenclature[V5]{$\hat{G}_{\tau \mid m}$}{The forecasted station charging load at time index $\tau$, assuming new user $n$ chooses charging option $m$, in kW.}
\nomenclature[V6]{$D_\tau$}{Station peak power up to time $\tau$, in kW.}
\nomenclature[V7]{$\zeta_i$}{Fixed charging price for existing user $i$ at the time index of the current optimization, in \$/{\rm kWh}.}
\nomenclature[V8]{$z^m$}{Charging tariff for new user $n$ who selects option $m$, in \$/kWh.}
\nomenclature[V9]{$z$}{Set of prices $(z^\text{sch}, z^\text{reg})$ to be presented to new user $n$, in \$/{\rm kWh}.}
    
\begin{document}

\title{Joint Price and Power MPC for Peak Power Reduction at Workplace EV Charging Stations\thanks{This work was supported in part by TotalEnergies S.E. Corresponding author: S. Bobick. T. Cambronne and S. Bobick contributed equally to this work.}}

\author{\IEEEauthorblockN{Thibaud Cambronne\IEEEauthorrefmark{1}\IEEEauthorrefmark{2}, Samuel Bobick\IEEEauthorrefmark{1}\IEEEauthorrefmark{3}, Wente Zeng\IEEEauthorrefmark{4}, Scott Moura\IEEEauthorrefmark{1}\IEEEauthorrefmark{2}}
\IEEEauthorblockA{\IEEEauthorrefmark{1}University of California, Berkeley, CA 94720\\}
\IEEEauthorblockA{\IEEEauthorrefmark{2}Department of Civil and Environmental Engineering}
\IEEEauthorblockA{\IEEEauthorrefmark{3}Department of Electrical Engineering and Computer Sciences}
\IEEEauthorblockA{\IEEEauthorrefmark{4}TotalEnergies S.E, Berkeley, CA 94704
\\ Emails: \{thibaudcambronne, samuelbobick, smoura\}@berkeley.edu, wente.zeng@totalenergies.com}}

\maketitle

\begin{abstract}

Demand charge, a utility fee based on an electricity customer's peak power consumption, often constitutes a significant portion of costs for commercial electric vehicle (EV) charging station operators. This paper explores control methods to reduce peak power consumption at workplace EV charging stations in a joint price and power optimization framework. We optimize a menu of price options to incentivize users to select controllable charging service. Using this framework, we propose a model predictive control approach to reduce both demand charge and overall operator costs. Through a Monte Carlo simulation, we find that our algorithm outperforms a state-of-the-art benchmark optimization strategy and can significantly reduce station operator costs.

\end{abstract}

% \begin{IEEEkeywords}
% Demand charge, dynamic pricing, electric vehicles, model predictive control, time series, optimization, smart charging, smart grid
% \end{IEEEkeywords}

% \renewcommand{\nomgroup}[1]{%
%   \ifthenelse{\equal{#1}{I}}{%
%     \item[\textit{Indices/Sets}]}{%
%   \ifthenelse{\equal{#1}{P}}{%
%     \item[\textit{Parameters}]}{%
%   \ifthenelse{\equal{#1}{V}}{%
%     \item[\textit{Variables}]}{}}}
% }

\printnomenclature

% \tableofcontents

\section{Introduction}

Smart electric vehicle (EV) charging at workplaces presents many key opportunities to contribute to transportation electrification and decarbonization. For example, although EV owners typically charge overnight, midday workplace charging capitalizes on cheap, renewable-laden electricity, thus realizing the full emissions reduction potential of EVs \cite{powell2022charging}. Also, workplace charging can alleviate inequities for EV owners without access to at-home chargers, such as those in multi-unit dwellings \cite{peterson2011addressing}. Furthermore, intelligent charging algorithms can help align charging with utility time-of-use (TOU) rate schedules and shave peak power consumption to maximize profit and incentivize charger expansion by private companies \cite{williams2014pricing}.

A major cost for some EV charging station operators is demand charge, a utility fee based on the highest power consumption during a billing cycle. Approximately 5 million commercial customers in the United States are estimated to face retail electricity tariffs with demand charges greater than \$15/kW, accounting for more than a quarter of the nation’s 18 million commercial customers \cite{mclaren2017identifying}. Demand charges are a significant component of commercial utility bills, generally accounting for 30\% to 70\% of total electricity costs \cite{mclaren2017identifying}.

Demand charge reduction through smart charging presents a challenging control problem. Even with smart charging technology, EV charging station arrivals are stochastic. On busy days, a single ill-timed arrival can push the station's load above the running peak, increasing demand charge.

Most prior work on demand charge mitigation for EV charging stations focuses on optimizing power schedules to maximize profit while accounting for EV arrival and departure times, energy demand constraints, demand charges, and TOU utility tariffs \cite{lee2020pricing, powell-transformer, kara-benefits, chaudhari2022learning, Yang2024}. Lee, Pang, and Low \cite{lee2020pricing} propose an offline pricing scheme that optimizes power delivery under fixed arrival, departure, and energy demand constraints. Their approach ensures cost recovery by incorporating facility demand charges, time-varying energy costs, and congestion costs. However, it does not account for stochastic variations in EV charging behavior. Yang et al. \cite{Yang2024} use ``block'' model predictive control to optimize power delivery for demand charge mitigation under uncertain future arrivals, departures, and energy demands. However, their approach does not consider how to structure pricing to encourage user behavior that mitigates demand charge.

Formulations in \cite{zeng2021, obeid2023learning, Ozturk2025} optimize both power delivery and user-facing prices. By incentivizing users with discounts if they allow their load to be controlled instead of uncontrolled, they are able to increase revenue. However, these formulations only control on-going charging sessions. At workplace stations, most charging sessions begin in the morning and continue until the workday ends. Consequently, control decisions made at 8 AM can influence peak power at 2 PM. Therefore, early indicators of a peak-inducing day should be incorporated into morning control decisions. An effective controller needs anticipatory ability.

To solve this problem, we propose two solutions that provide anticipatory abilities: (i) convex reformulations of the demand charge constraint, and (ii) a forecast-enabled model predictive control (MPC) approach. This enables the following novel contributions to the existing literature:

    \begin{enumerate}
        \item We propose a MPC method that anticipates demand spikes using time series forecasting. 
        \item We increase the load flexibility available to the controller by leveraging dynamic pricing to incentivize users to choose a controllable charging service.
        \item We demonstrate that our approach outperforms the benchmark optimization algorithm from \cite{zeng2021} and other non-MPC reformulations in reducing both demand charge and overall operator costs.
        % \item We demonstrate that jointly optimizing user-facing prices to incentivize user behavior along with power delivery is more effective for both peak power and energy cost reduction than optimizing for power delivery only.
        % \item There is a strong relationship between forecast error and MPC controller performance.
        % \item Using pricing to incentivize choosing scheduled, which gives the controller flexibility for demand charge reduction
        % \item Spreads the cost burden of demand charge across users more equitably
    \end{enumerate}

\section{Benchmark Optimization Problem}
\label{sec:baseline}

In this section, we present a benchmark station-level optimization problem that is solved when a new user arrives at the charging station. Our formulation closely follows the formulation presented in \cite{zeng2021}, with some modifications.

Upon arrival at the charging station, user $i$ inputs their required energy $\E_i^\req$ and expected departure time from the charging station $T_i^{\text{end}}$. Given this information, the user is presented with two charging prices, $z_\text{sch}$ and $z_\text{reg}$, generated by a price and power optimization algorithm. Given these prices, the user can make one of three choices:

\begin{enumerate}
    \item \reg{}: the battery charges at maximum power until it tops off or unplugs. Importantly, this load is uncontrollable.
    \item \sch{}: the user's energy required is guaranteed to be provided by their indicated time of departure. The power delivery is scheduled with a control algorithm.
    \item \lea{}: The user can leave if they consider the prices to be too high, opting for nearby chargers or a parking spot without a charger. Without this option, the optimal service price that maximizes profit is infinite, which is non-sensical.
\end{enumerate}

The control objective is to maximize the expected profit from the station by choosing the \reg{} and \sch{} prices presented to the new user, along with the power profiles for all users who selected the \sch{} charging option.

The benchmark solution is a conventional non-anticipatory optimization that involves solving (\ref{eqn:obj}) directly. We assume that users choose one of the 3 charging options based solely on the price menu. We denote $\Pr{m \mid z}$ to be the probability that new user $n$ chooses option $m$ when presented with the set of prices $z$. We model these probabilities with a two-step discrete choice model, described in Appendix \ref{app:dcm}. Due to the non-convexity of the discrete choice model \eqref{eqn:dcm1}--\eqref{eqn:dcm2}, optimization problem \eqref{eqn:obj} is difficult to solve numerically. As such, we find an approximate solution to (\ref{eqn:obj}) by searching over a grid of candidate prices $z_\text{candidate} = (z_\text{sch}, z_\text{reg})$. At each point in the grid, we solve $\min_{p} \mathsf{E}[{ f(p, m) \mid z_\text{candidate}}]$, which is a linear program.

\subsection{Objective Function}

The joint price and power objective, defined as a function of prices $z$ and power profiles $p$, is expressed as an expectation over new user's choice $m$. We formulate the resulting optimization problem as follows:

\begin{align}
    \min_{z,p} ~~ &\mathsf{E}[f(p, m) \mid z] \label{eqn:obj}\\ 
    = & \  \Pr{m=\sch{} \mid z} f^{\text{sch}}(z, p) \nonumber \\
    + & \ \Pr{m=\reg{} \mid z} f^{\text{reg}}(z, p) \nonumber \\
    + & \ \Pr{m=\lea{} \mid z} f^{\text{leave}} \nonumber\\
    \sto \qquad 
    &\text{Demand charge constraints \eqref{eqn:dc1}--\eqref{eqn:dc2}}, \nonumber \\ &\text{Energy constraints \eqref{eqn:power_bound}--\eqref{eqn:energy-req}}.\nonumber 
\end{align}

Here, $f^{\text{sch}}$, $f^{\text{reg}}$, and $f^{\text{leave}}$ represent the station operator's cost functions associated with each potential choice made by the new user. The optimization horizon runs from current time index $T^{\text{start}}$ up to but not including $T^{\text{end}}$, the latest indicated time of departure across all users $i \in \mathcal{I}$.

\subsection{Demand Charge Constraints}

The running peak power is tracked over a monthly billing cycle, consistent with utility practice. The billing cycle starts at $\tau = 0$. $G_\tau$ is the station load at time $\tau$:

\begin{align}
    G_\tau &= \sum_{i \in A_{\text{sch}}} p_{i,\tau}^\text{sch} + \sum_{j \in A_{\text{reg}}} p^{\text{max}}\mathsf{1} \{ \tau < T_j^\text{end}\} \quad \forall \tau \geq 0. \label{eqn:dc1}
\end{align}

The running peak power $D_{\tau}$ is defined as follows:

\begin{align}
    D_{0} &= 0,\\
    D_{\tau+1} &= \max\{G_{\tau + 1}, D_\tau\} \quad \forall \tau \geq 0 \label{eqn:dc2}.
\end{align}

\subsection{Scheduled Cost Function}    
Provided with new user $n$'s energy required $E_n^\text{req}$ and time index of departure $T_n^\text{end}$, the cost function assuming that user $n$ chooses the \sch{} charging option is given by

\begin{align}
    f^{\text{sch}}\left(z, p\right)
    &= \sum_{\tau=T^\text{start}}^{T_n^\text{end} - 1} \left(c_\tau - z^{\text{sch}}\right) p_{n,\tau}^{\text{sch}} \Delta t 
    \label{eqn:sch1} \\
    &+ \sum_{i \in \mathcal{A}_{\text{sch}}} \left[ \sum_{\tau=T^\text{start}}^{T^\text{end}_i - 1} \left(c_\tau - \zeta_i\right) p_{i,\tau}^{\text{sch}} \Delta t \right] 
    \nonumber  \\
    &+ \sum_{j \in \mathcal{A}_{\text{reg}}} \left[ \sum_{\tau=T^\text{start}}^{T^\text{end}_j - 1} \left(c_\tau - \zeta_j\right) p^{\text{max}} \Delta t \right] 
    \nonumber  \\
    &+ c_D \left(D_{T^{\text{end}} - 1} - D_{T^\text{start} - 1}\right). \nonumber 
\end{align}

The final term of the objective, $c_D \left(D_{T^{\text{end}} - 1} - D_{T^\text{start} - 1}\right)$, represents the cost associated with the \textit{increase in monthly peak power} over the optimization horizon \([T^\text{start}, T^\text{end})\). Note that, as indicated by \eqref{eqn:dc2}, this term cannot be negative and only becomes positive when the monthly peak power increases over the optimization horizon.

% \blue{Additionally, the cost for this increase is assigned exclusively to the new user \(n\), even though other existing users may also be charging during the peak period.} \red{Delete the last sentence probably, not a focus of the paper}

\subsection{Regular Cost Function}
Provided with new user $n$'s energy required $E_n^\text{req}$ and the time index $T_n^\text{end}$ when the user's battery would be fully charged when charging at $p^\text{max}$, the cost function assuming that user $n$ chooses the \reg{} charging option is given by

\begin{align}
    f^{\text{reg}}\left(z, p\right)
    &= \sum_{\tau=T^\text{start}}^{T_n^\text{end} - 1} \left(c_\tau - z^{\text{reg}}\right) p^{\text{max}} \Delta t 
    \label{eqn:reg1} \\
    &+ \sum_{i \in \mathcal{A}_{\text{sch}}} \left[ \sum_{\tau=T^\text{start}}^{T^\text{end}_i - 1} \left(c_\tau - \zeta_i\right) p_{i,\tau}^{\text{sch}} \Delta t \right] 
    \nonumber \\
    &+ \sum_{j \in \mathcal{A}_{\text{reg}}} \left[ \sum_{\tau=T^\text{start}}^{T^\text{end}_j - 1} \left(c_\tau - \zeta_j\right) p^{\text{max}} \Delta t \right] 
    \nonumber \\
    &+ c_D \left(D_{T_{\text{end}} - 1} - D_{T_\text{start} - 1}\right)
    \nonumber.
\end{align}

% Analogous to the scheduled case, \eqref{eqn:reg4} captures the cost associated with the increase in peak power over optimization horizon [$T^\text{start}$, $T^\text{end}$).

% Note that \eqref{eqn:reg2}--\eqref{eqn:reg4} are the same as \eqref{eqn:sch2}--\eqref{eqn:sch4}, but the objective value may be different since user $n$'s power cannot be controlled here.

\subsection{Leave Cost Function}

If the user chooses to leave, the station operator incurs neither cost nor benefit. Thus,  $f^{\text{leave}} = 0$.

\subsection{Energy Constraints}

Constraint (\ref{eqn:power_bound}) ensures that the charging power lies within the charger rating, (\ref{eqn:energy_state_evolution}) tracks the battery state of charge $e_{i, \tau}$, and (\ref{eqn:energy-req}) ensures that the chosen power profiles satisfy each user $ i \in \mathcal{I}$'s energy requirements $E_i^\text{req}$:

\begin{align}
    0  
    &\leq p_{i, \tau}^{\text{sch}} \leq p^{\text{max}} 
    && \forall i \in \mathcal{I}, \; \tau \in [T^{\text{start}}, T_i^{\text{end}}),
    \label{eqn:power_bound} \\
    e_{i, \tau} 
    &= e_{i, \tau - 1} + \Delta t \eta p_{i,\tau} 
    && \forall i \in \mathcal{I}, \; \tau \in [T^{\text{start}}, T_i^{\text{end}}), 
    \label{eqn:energy_state_evolution} \\
     E_i^{\text{req}} 
    &\leq e_{i, T_i^{\text{end}}} 
    && \forall i \in \mathcal{I}.
    \label{eqn:energy-req}
\end{align}

\section{Controllers for Improved Peak Power Management}\label{sec:controllers}

In this section, we propose several modifications and improvements to optimization problem (\ref{eqn:obj}) which aim to reduce monthly demand charge. Section \ref{sec:threshold} applies a hard constraint to peak power. Section \ref{sec:variable} gives extra weight to the demand charge term when the current power profile is on the verge of exceeding peak power. Section \ref{sec:timeseries-mpc} uses MPC to give the controller an anticipatory ability.

\subsection{Iterative Hard Thresholding}
\label{sec:threshold}

In this approach, we add a hard constraint $M$ on station load $G_\tau$ to find the feasible solution that results in the lowest increase in peak power. That is, we solve (\ref{eqn:obj}) with the additional constraint

\begin{equation}\label{eqn:threshold}
    G_\tau \leq M \quad \forall \tau \in [T^\text{start}, T^\text{end}).
\end{equation}

$M$ is initialized at the running peak in the billing cycle so far, $D_{T^\text{start} - 1}$. If \eqref{eqn:obj} with constraint \eqref{eqn:threshold} is infeasible, we iteratively loosen the constraint by increasing $M$ by some step size $\epsilon > 0$, and re-solve until we find a feasible solution.

\subsection{Softplus Demand Charge Penalty}
\label{sec:variable}

\begin{figure}
    \centering
    \includegraphics[width=\linewidth]{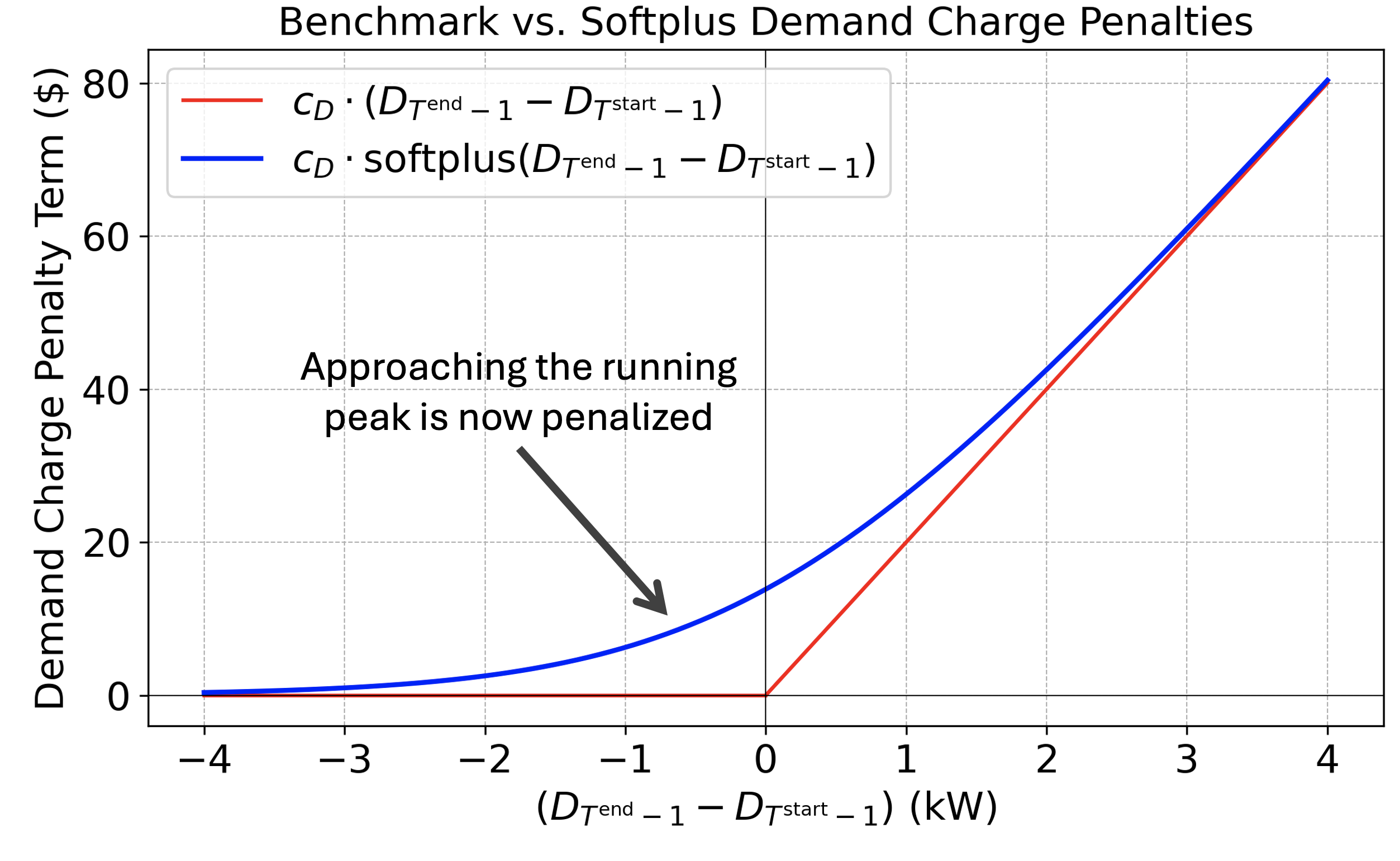}
    \caption{The softplus function adds a penalty when the peak of the optimized station power is approaching the running peak (as $x \xrightarrow{}0^{-}$), and converges to the original penalty as the station power exceeds the running peak (as $x \xrightarrow{}+\infty$).}
    \label{fig:softplus.png}
\end{figure}

Approaching the running peak is undesirable because any future arrival is likely to increase peak power. Thus, we penalize this case by applying the convex softplus function, $\text{softplus}(x) = \log(1 + e^x)$, to the increase in peak power over the billing cycle, as illustrated in Fig. \ref{fig:softplus.png}. That is, we replace the demand charge terms in \eqref{eqn:sch1} and \eqref{eqn:reg1} with 

\begin{equation}
    c_D \text{softplus} \left(D_{T^{\text{end}} - 1} - D_{T^\text{start} - 1}\right).
\end{equation}

\subsection{Model Predictive Control}
\label{sec:timeseries-mpc}

Next, we incorporate a time series forecast into the demand charge term to anticipate arrivals. Let $\{ \hat{G}_{T_\text{start} \mid m}, ..., \hat{G}_{T_\text{start} + l \mid m} \}$ represent the forecasted station power time series for the next $l$ timesteps, given that new user $n$ chooses charging option $m$. Then, we can replace the demand charge terms in \eqref{eqn:sch1} and \eqref{eqn:reg1} with

\begin{equation}
    c_D \left( \max \{ \hat{G}_{T^\text{start} \mid m}, ..., \hat{G}_{T^\text{start} + l \mid m}, D_{T^\text{start} - 1} \} - D_{T^\text{start} - 1} \right).
\end{equation}

We test the MPC algorithm with 3 types of forecasters:

\begin{enumerate}
    \item A naive model that assumes no additional arrivals, but uses, as a forecast, the sum of the optimal power profile from the \textit{previous} optimization with new user $n$'s power profile, assuming they chose the \reg{} option; 
    \item A linear model trained with elastic net regression \cite{zou2005regularization};
    \item XGBoost \cite{chen2016xgboost}.
\end{enumerate}

Forecasting models 2) and 3) use features such as recent historical power profiles, planned power profiles from the last optimization, number of sessions, and encodings of time of day, week, year, and workday. Model hyperparameters are tuned with a train-validation-test split of empirical station usage data. More details on feature engineering and forecast implementation can be found in Appendix \ref{app:model-details}. 

\section{Numerical Study}

\begin{table*}
    \centering
    \caption{Profit Comparison Across Control Schemes}
    \setlength{\tabcolsep}{1.1pt} % Reduce column spacing for better fit
    \renewcommand{\arraystretch}{1.1} % Adjusts row height
    \label{tab:profit-comparison}
    \footnotesize % Use smaller font to fit IEEE two-column format
    \begin{tabular}{m{2cm} | >{\centering\arraybackslash}m{1.5cm} >{\centering\arraybackslash}m{1.5cm} >{\centering\arraybackslash}m{1.5cm} >{\centering\arraybackslash}m{1.5cm} >{\centering\arraybackslash}m{1.5cm} | >{\centering\arraybackslash}m{1.5cm} >{\centering\arraybackslash}m{1.5cm} >{\centering\arraybackslash}m{1.5cm}}
        \hline \hline
        \multirow{2}{=}{\centering \textbf{Control Scheme}} & \multicolumn{5}{c|}{\textbf{Mean Cost or Revenue (\$)}} & \multicolumn{3}{c}{\textbf{Change from Benchmark (\%)}} \\
        \cline{2-9}
        & \textbf{Demand Charge} & \textbf{TOU} & \textbf{Revenue} & \textbf{Cost} & \textbf{Profit} & \rule{0pt}{12pt} \textbf{Demand Charge} & \textbf{TOU} & \textbf{Cost} \\
        \hline
        Benchmark & 624 & 1,984 & 4,309 & 2,608 & 1,702 & 0.00 & 0.00 & 0.00 \\
        Threshold & 628 & 1,987 & 4,331 & 2,615 & 1,716 & -0.71 & 0.16 & +0.29 \\
        Softplus & 609 & 2,017 & \textbf{4,446} & 2,626 & 1,819 & -2.34 & 1.67 & +0.71 \\
        MPC (Naive) & \textbf{519} & 1,970 & 4,253 & \textbf{2,488} & 1,765 & \textbf{-16.90} & -0.71 & \textbf{-4.58} \\
        MPC (Linear) & 521 & 1,970 & 4,362 & 2,491 & 1,870 & -16.50 & -0.68 & -4.47 \\
        MPC (XGBoost) & 533 & \textbf{1,956} & 4,434 & 2,489 & \textbf{1,944} & -14.57 & \textbf{-1.39} & -4.55 \\
        \hline \hline
    \end{tabular}
\end{table*}

\subsection{Simulation Overview}

We perform a Monte Carlo simulation\footnote{Simulation code available at \url{https://github.com/samuelBobick/StationLevelPowerForecasting}.} to quantitatively validate the performance of the control algorithms presented in the previous section. We replay all EV charging sessions sessions on the Smart Learning Pilot for EV Charging Stations (SlrpEV) platform from 2023, a total of 2274 charging sessions \cite{Ozturk2025}. We assume a demand charge of \$20/kW \cite{mclaren2017identifying}, representing 24\% of simulated total station operator costs under the benchmark algorithm. We use the Pacific Gas \& Electric  Commercial Business Electric Vehicle June 2023 rates for the TOU tariff structure \cite{pge_tariff}. Note that the computational complexity of the algorithms proposed in Section \ref{sec:controllers} is not prohibitive; all algorithms run within in a fraction of a second on a standard laptop. Further details on the simulation setup can be found in Appendix \ref{app:sim-setup}.

\subsection{Results and Discussion}

Table \ref{tab:profit-comparison} presents the Monte Carlo simulation results, obtained by running each of the 12 months 10 times.

In particular, all MPC schemes successfully decrease demand charge by approximately 15\% from the benchmark, while also decreasing TOU costs by approximately 1\%, resulting in an approximately 4.5\% average total cost reduction. The two non-MPC controllers fail to significantly reduce the total costs.

\begin{figure}
    \centering
    \includegraphics[width=\linewidth]{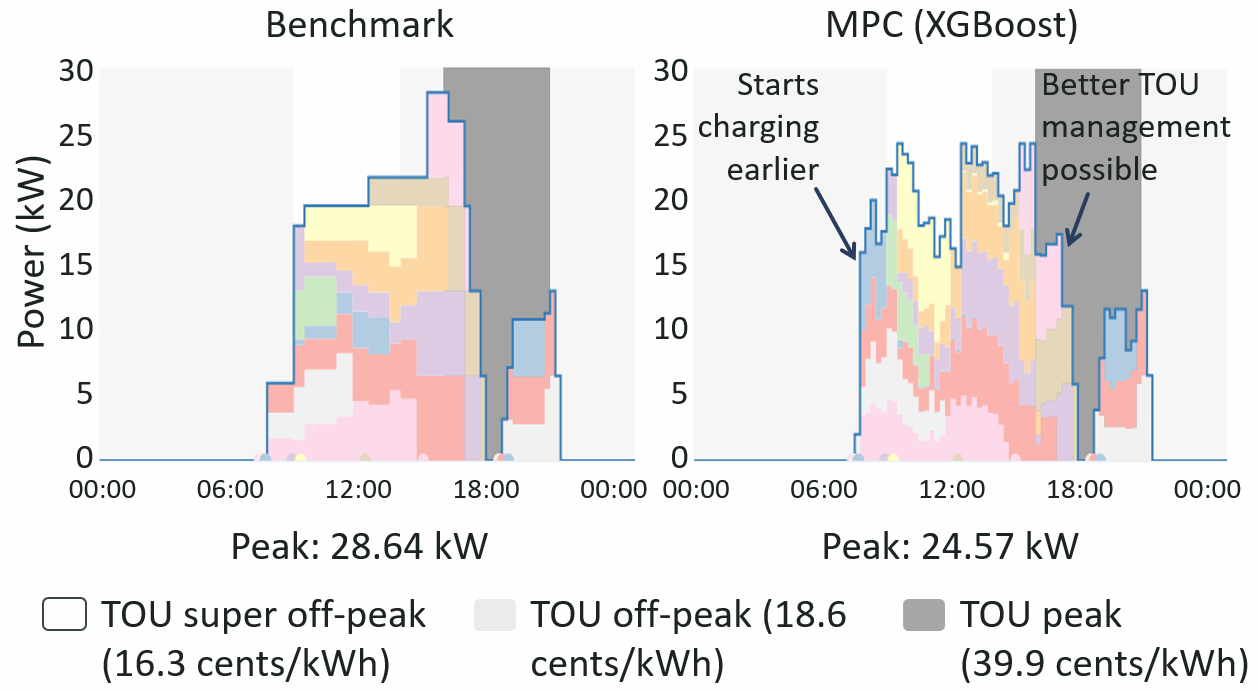}
    \caption{Example of control actions for benchmark (left) and MPC (right) on September 6th, 2023. For these two examples, the users all arrived at the same time, with the same requirements and all chose \sch{}. Each color represents a single user's power profile. The MPC algorithm shows better repartition of the load throughout the off peak hours and achieves a lower peak (24.57 kW) compared to the benchmark solution (28.64 kW), while also reducing the energy consumed during peak TOU hours.}
    \label{fig:control-boxplot}
\end{figure}

Figure \ref{fig:control-boxplot} demonstrates how MPC improves peak power management. As the day's first sessions begin, the benchmark controller waits to fulfill energy demand until the cheapest energy (super off-peak) is available. On the other hand, the MPC controller anticipates a busy day, and starts charging earlier so that there is less demand during the peak power event. In fact, by aggressively charging early in the day, the MPC algorithm is even able to partially avoid peak TOU pricing periods.

\begin{figure}
    \centering
    \includegraphics[width=\linewidth]{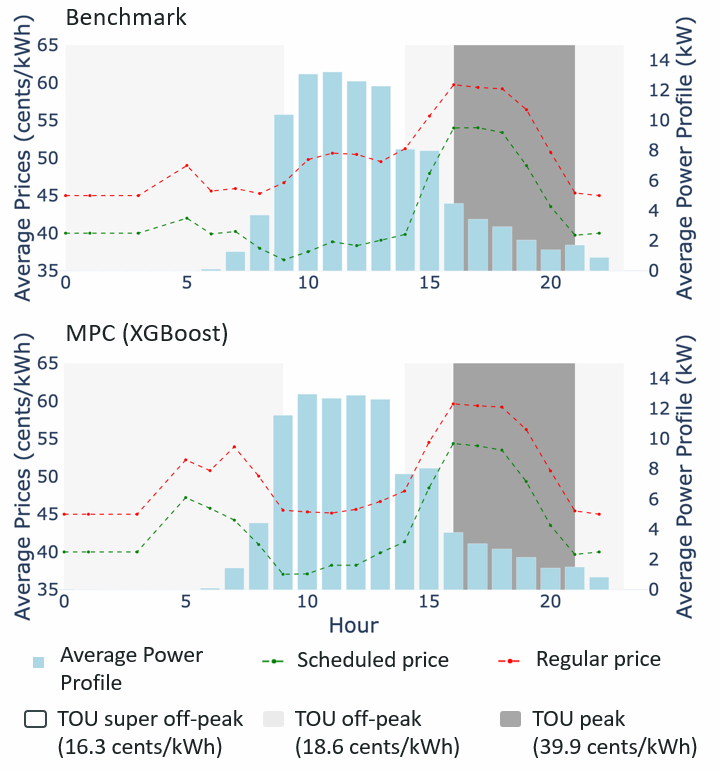}
    \caption{Example of pricing strategy for benchmark (top) and MPC (bottom) solutions, overlaid with the average station power profile.}
    \label{fig:pricing-plot}
\end{figure}

Figure \ref{fig:pricing-plot} illustrates how the MPC controller leverages pricing to encourage users to select \sch{} when demand charge mitigation is pertinent. Both strategies raise prices before and during peak TOU hours when energy costs double. However, the MPC algorithm sets high prices early in the morning while significantly discounting \sch{} between 6--8 AM, prioritizing a base of controllable \sch{} sessions for the day. As a result, the MPC controller is able to maintain lower daytime prices, particularly for \reg{} sessions, whereas the benchmark controller must raise prices during peak load hours to compensate for unanticipated sessions.

% \subsection{The Value of Joint Price and Power Optimization}
% \blue{TODO when the results finish running.}

\subsection{Forecast Performance vs. MPC Peak Power Reduction}
\label{sec:forecast-analysis}

Table \ref{tab:rmse} compares forecast errors to simulated mean peak power. The training RMSE closely correlates with the RMSE of predictions generated during the simulation. However, the naive forecast achieves the best peak power reduction despite a much higher training RMSE and a higher simulation RMSE. This suggests that minimizing RMSE alone does not strongly correspond to improved MPC performance. 

\begin{table}
    \centering
    \caption{The Value of Forecast Accuracy in MPC}
    \label{tab:rmse}
    \begin{tabular}{p{2cm} p{1.5cm} p{1.5cm} p{1.5cm}}
        \hline
        \textbf{Control Scheme} & 
        \textbf{Training RMSE (kW)} &
        \textbf{Simulation RMSE (kW)} &
        \textbf{Mean Peak Power (kW)} \\
        \hline
        MPC (Naive) & 11.36 & 7.46 & \textbf{25.93} \\
        MPC (Linear) & 6.25 & \textbf{4.13} & 26.05 \\
        MPC (XGBoost) & \textbf{5.94} & 4.31 & 26.65 \\
        \hline
    \end{tabular}
\end{table}

One possible explanation for this is that the linear and XGBoost models were trained on data that does not exactly match the distribution of simulation data. Specifically, the training data contains charging sessions which were controlled with an optimization algorithm similar to the benchmark algorithm. On the other hand, the naive model is based on the previous output of the MPC controller, making it a perfect forecast if no new vehicles arrive on that day and a good approximation even if a few do. Therefore, retraining the models on data generated by the MPC controller would better match forecasts with their role in the optimization formulation. But this is challenging, as forecasts depend on MPC outputs cannot be determined without the forecasts, creating a self-referential problem.

Our test case, the SlrpEV station, has eight chargers. Due to its small size, station loads are stepwise and irregular, making forecasting more challenging. For a larger set of EV chargers, the Law of Large Numbers would smooth load curves, increasing periodicity and improving forecast accuracy. Consequently, the MPC approach may perform even better at larger stations or across a collection of aggregated station loads.

\section{Conclusions and Limitations}

In this paper, we present several methods to decrease demand charge at workplace EV charging stations in a joint price and power optimization scheme and use a Monte Carlo simulation to validate their performance against a state-of-the-art benchmark optimization algorithm. The MPC implementation detailed in Section \ref{sec:timeseries-mpc} decreases overall station operator costs by 4.6\% and demand charge by 16.9\%.

% In the benchmark solution, the user whose arrival pushes the station load above its peak bears a disproportionate share of the demand charge costs, even though other users are already contributing to the peak. Our MPC approach addresses this issue by distributing the cost burden of demand charge expenses across all users whose usage is \textit{forecasted} to contribute to an increase in peak power, ensuring a more fair allocation. A detailed analysis of how demand charge costs are allocated fairly across users via online pricing presents a potential avenue for future research.

In Section \ref{sec:forecast-analysis}, we find that minimizing forecast RMSE alone does not strongly correlate with improved MPC performance. Li, Ju, and Wang \cite{li2024quantifying} have a similar finding in the context of building energy management. Training the models on data generated by the MPC algorithm would be more accurate, but this introduces a self-referential problem that requires future work. 
% The authors find that forecasting errors have an asymmetric impact on MPC performance when demand charge is part of the objective. Further work is required to develop forecasting error metrics and techniques that integrate well with demand charge mitigation controllers. 

When simulating SlrpEV user sessions, we randomly select between \reg{} and \sch{} choices with probabilities specified by the discrete choice model. While we account for \lea{} in the optimization problem presented in \eqref{eqn:obj}, in the Monte Carlo simulation we are limited in the fact that users have already decided not to \lea{}. This limitation arises because the SlrpEV dataset only contains information about users who chose to charge. Thus, profit and revenue should be interpreted with caution. Ultimately, station operators optimize for profit, so further work is needed to incorporate \lea{} behavior into the Monte Carlo simulation. 

Joint price and power optimization is both technically feasible and shown to have high user acceptance. Indeed, cloud-connected Level 2 power management is already offered by ChargePoint and EVgo, albeit without \textit{joint} price and power optimization \cite{chargepoint, evgo}. Moreover, prior work uses real-world experiments to show that significant proportions of customers are willing adopt \sch{} charging in exchange for discounts \cite{obeid2023learning, Ozturk2025}. One shortcoming of our dataset is the fact that the SlrpEV station only has eight chargers. Future work should examine the scalability of joint price and power optimization considering peak power mitigation over tens, hundreds, or thousands of EV chargers by a single operator across many sites.

Ultimately, we find that MPC significantly decreases charging station operating costs, particularly demand charge. These savings can either boost operator profits or be used to offer lower prices to customers. Enhancing behavioral simulation, testing our algorithms for control of large aggregate loads, and experimentally deploying our control schemes with real human behavior as in \cite{Ozturk2025, lee-adaptive} are key steps toward further improvement of joint price and power control for demand charge mitigation at workplace EV charging stations.

\appendices
\section{Behavioral Model Formulation}
\renewcommand{\theequation}{A.\arabic{equation}} % Prefix equations with "A."
\setcounter{equation}{0} % Reset equation counter

\label{app:dcm}

Our behavioral model describes the probability that user $n$ selects charging choice $m$ when presented with prices $z$. Since $z$ represents prices per unit of energy (\$/kWh), we multiply it by the maximum charging power $p^\text{max}$ to convert it to a price per unit of time (\$/hr). This transformation ensures consistency between the optimization framework in Section \ref{sec:baseline} and the utility model in \cite{Ozturk2025}.

\begin{equation}
z_\text{reg}'= z^\text{reg}p^\text{max}, \quad
z_\text{sch}'= z^\text{sch}p^\text{max}
\end{equation}

To estimate the choice probabilities, we use a two-step discrete choice model, using utilities $U$ estimated in \cite{Ozturk2025}.

\begin{align}
        U_\text{reg} &= 0.341 - 0.0184\frac{(z'_\text{reg} - z'_\text{sch})}{2} \\
        U_\text{sch} &= 0.0184\frac{(z'_\text{reg} - z'_\text{sch})}{2} \\
        U_\text{leave} &= -1 + 0.005\frac{(z'_\text{reg} + z'_\text{sch})}{2} 
\end{align}

First, we calculate the probability that the user chooses the \lea{} option:

\begin{align} 
    \Pr{m = \lea{} \mid z}
    &= \frac{e^{U_\text{leave}}}{e^{U_\text{sch}} + e^{U_\text{reg}} + e^{U_\text{leave}}} .\label{eqn:dcm1}
\end{align}

Then, we calculate the probability that the user chooses \sch{}, given that they do not choose the \lea{} option:

\begin{align} 
    \text{Pr}(m &= \sch{} \mid z, m \neq \lea{}) \\
    &= \frac{e^{U_\text{sch}}}{e^{U_\text{sch}} + e^{U_\text{reg}}}
    (1 - \text{Pr}(m = \lea{} \mid z)).
\end{align}

Likewise, for \reg{},

\begin{align} 
    \text{Pr}(m &= \reg{} \mid z, m \neq \lea{}) \\
    &= \frac{e^{U_\text{reg}}}{e^{U_\text{sch}} + e^{U_\text{reg}}}
    (1 - \text{Pr}(m = \lea{} \mid z)).  \label{eqn:dcm2}
\end{align}

\section{Time Series Forecasting Details}
\label{app:model-details}

Our model forecasts the station power time series for the next $l$ timesteps. It takes the following as features\footnote{More external features (such as weather, special events, etc.) could be further introduced to improve the prediction accuracy. However, we don't consider them in this paper, as the focus of this work is the MPC formulation.}:

\begin{enumerate}
    \item The $k$ most recent aggregate power values: $G_{T_\text{start} - k}, G_{T_\text{start} - k + 1}, ... G_{T_\text{start} - 1}$;
    \item The planned station power profile for the next $l$ timesteps for all active users in $\mathcal{I}$, considering that user $n$ chooses \reg, and given power profiles $\bar{p}$ generated from the previous optimization (triggered when user $n-1$ arrived) for users in $\mathcal{A}_\text{sch} \cup \mathcal{A}_\text{reg}$:
    $G_{T_\text{start} | \reg{}, \bar{p}}, ... G_{T_\text{start} + l | \reg{}, \bar{p}}$;
    \item The number of active sessions at the charging station, including new user $n$: $|\mathcal{I}|$;
    \item The total number of working chargers at the station at time index $T^\text{start}$;
    \item A sinusoidal positional encoding of the time of day, time of week, and time of year \cite{vaswani2017attention};
    \item A boolean variable that evaluates to 1 if the time index $T^\text{start} + 24$ hours falls on a workday.
\end{enumerate}

% We generate our forecast assuming new user $n$ chooses the \reg{} charging option; i.e. our forecaster predicts $\hat{G}_{\tau \mid \reg{}}$.

Let \(\psi_\tau\) represent the forecast of the station power profile, made outside of the control loop. At each solver iteration, to calculate $\hat{G}_{\tau \mid \reg{}}$ and $\hat{G}_{\tau \mid \sch{}}$ for time indices $\tau \in [T^\text{start}, T^{\text{end}})$, we can simply start from the original forecast $\psi_\tau$ and add and subtract power profiles of the new control input $p$. Let $\bar{p}^\text{sch}_i$ represent the optimal scheduled power profile for user $i$ found in the \textit{previous} optimization triggered by user $n-1$. Then, we update our forecast as follows:

\begin{equation}
    \label{eqn:update1}
    \hat{G}_{\tau \mid \reg{}} = \psi_\tau + \sum_{i \in \mathcal{A}_{\text{sch}}} \left(  p_{i, \tau}^\text{sch}  - \bar{p}_{i, \tau}^\text{sch} \right)
\end{equation}

\begin{equation}
    \hat{G}_{\tau \mid \sch{}} = \psi_\tau + (p_{n, \tau}^\text{sch}  - p_{n, \tau}^\text{reg}) + \sum_{i \in \mathcal{A}_{\text{sch}}} \left(  p_{i, \tau}^\text{sch}  - \bar{p}_{i, \tau}^\text{sch}\right) \label{eqn:update2}
\end{equation}

For each new user, the forecast is only executed once outside of the control loop, and is arithmetically updated to the new control actions at each grid search iteration of the solver. Compared to executing the forecast inside the control loop, our single forecast approach allows us to:
 \begin{enumerate}
    \item Improve forecaster performance by using non-convex forecasters versus being limited to convex models, or having to increase computational time by using non-convex optimization;
    \item Improve the forecast reliability by making predictions only on controlled historical power profiles versus candidate power profiles tested by the solver, which may not match the distribution of data that the forecaster was trained on;
    \item Reduce computational time by making a single prediction instead of one per solver iteration.
 \end{enumerate}

We employ separate models for workdays and non-workdays (i.e., weekends and holidays). The model is trained on the SlrpEV dataset \cite{Ozturk2025}.
Note that we post-process the forecast by clipping it to eliminate unrealistic forecasts:

\begin{align}
    \psi_\tau &\geq p_{n, \tau}^\text{reg} + \sum_{i \in \mathcal{A}_{\text{sch}}} \bar{p}_{i, \tau}^\text{sch} +
    \sum_{i \in \mathcal{A}_{\text{reg}}} \bar{p}_{i, \tau}^\text{reg}
    \label{eqn:update3} \\
    &\forall \tau \in [T^\text{start}, T^{\text{end}}). \nonumber
\end{align}

\section{Simulation Setup}
\label{app:sim-setup}
The SlrpEV dataset used to derive our simulation has a mix of \reg{} and \sch{} charging sessions. As we re-generate user choices using the discrete choice model, we make several assumptions:

\begin{enumerate}
    \item Users have already made the decision to charge. That is, they cannot \lea{}. They only choose between \reg{} and \sch{} options.
    \item The energy demand for charging sessions that were in reality \reg{} but are randomly simulated to be \sch{} have an energy demand equal to 57\% of the energy delivered in the original charging session, corresponding to the average in the SlrpEV dataset.
    % We choose 57\% because in the SlrpEV dataset on average the $E^\text{req}$ for \sch{} sessions is 57\% of the energy consumed from a \reg{} session of the same length.
    \item We assume we know that user $n$'s time of departure is fixed and known. 
\end{enumerate}

Additional parameter settings are listed in Table \ref{tab:parameters}.

\begin{table}[H]
    \centering
    \renewcommand{\arraystretch}{1.1}
    \setlength{\tabcolsep}{6pt} % Adjust column spacing for readability
    \caption{Simulation Parameters}
    \label{tab:parameters}
    \footnotesize % Use smaller font to fit IEEE two-column format
    \begin{tabular}{l l p{4cm}}
        \hline
        \textbf{Parameter} & \textbf{Value} & \textbf{Description} \\
        \hline
        $\Delta t$ & 0.25 hours & Length of each time step \\
        $p^{\text{max}}$ & 6.6 kW & Maximum charging power \\
        $\eta$ & 1 & Charger efficiency \\
        $\epsilon$ & 1 kW & Threshold increment for solution \ref{sec:threshold} \\
        $k$ & 96 & Number of past time steps to use as features for the forecaster \\
        $l$ & 32 & Number of time steps ahead that the forecaster predicts \\
        \hline
    \end{tabular}
\end{table}

\FloatBarrier % Prevent floats from moving past this point

\bibliographystyle{IEEEtran}
\balance
\bibliography{export.bib}

\newpage

\end{document}